\documentclass[lettersize,journal]{IEEEtran}
\usepackage{amsmath,amsfonts}
\usepackage{algorithmic}
\usepackage{algorithm}
\usepackage{array}
\usepackage[caption=false,font=normalsize,labelfont=sf,textfont=sf]{subfig}
\usepackage{textcomp}
\usepackage{stfloats}
\usepackage{url}
\usepackage{verbatim}
\usepackage{graphicx}
\usepackage{multirow}
\usepackage{todonotes}
\usepackage{xcolor}
\usepackage[switch]{lineno}
\usepackage{cite}
\newcommand{\mname}[0]{\mbox{{LightReSeg}}}

\begin{document}

\title{Light-weight Retinal Layer Segmentation with Global Reasoning}
    
\author{Xiang He, Weiye Song, Yiming Wang, Fabio Poiesi, Ji Yi, Manishi Desai, Quanqing Xu, Kongzheng Yang, and Yi Wan
\thanks{This work was supported in part by the National Natural Science Foundation of China under Grant 51975336, and 62205181, in part by the Natural Science Foundation of Shandong Province under Grant ZR2022QF017, in part by the Natural Science Outstanding Youth Fund of Shandong Province under Grant 2023HWYQ-023, in part by the Taishan Scholar Foundation of Shandong Province under Grant tsqn202211038, in part by the NIH under Grant R01NS108464, in part by the Key Technology Research and Development Program of Shandong Province under Grant 2020JMRH0202 and 2022CXGC020701, in part by the Shandong Province New Old Energy Conversion Major Industrial Tackling Projects under Grant 2021-13, in part by the Key Research and Development Project of Jining City under Grant 2021DZP005, and in part by the Shandong University Education Teaching Reform Research Project under Grant 2022Y133, 2022Y124, and 2022Y312. (Corresponding author: Yi Wan.)}
\thanks{Xiang He is with the School of Mechanical Engineering, and also with the Joint SDU-NTU Centre for Artificial Intelligence Research (C-FAIR), Shandong University, Jinan, Shandong, China (e-mail: he\_xiang@mail.sdu.edu.cn).}
\thanks{Weiye Song, Quanqing Xu, Kongzheng Yang, and Yi Wan are with the School of Mechanical Engineering, Shandong University, Jinan 250061, Shandong, China (e-mail:  songweiye@sdu.edu.cn; 202214333@mail.sdu.edu.cn; 202234432@mail.sdu.edu.cn; wanyi@sdu.edu.cn).}
\thanks{Yiming Wang, Fabio Poiesi are with the Fondazione Bruno Kessler, Via Sommarive 18, Povo, TN 38123, Italy (e-mail: ywang@fbk.eu; poiesi@fbk.eu).}
\thanks{Ji Yi is with the Department of Biomedical Engineering and the Department of Ophthalmology, Johns Hopkins University, Baltimore, Maryland 21231, USA (e-mail: jiyi@jhu.edu).}
\thanks{Manishi Desai is with the Department of Ophthalmology, Boston University School of Medicine, Boston Medical Center, Boston 02118, USA (e-mail: madesai@bu.edu).}}

\maketitle

\begin{abstract}
Automatic retinal layer segmentation with medical images, such as optical coherence tomography (OCT) images, serves as an important tool for diagnosing ophthalmic diseases. However, it is challenging to achieve accurate segmentation due to low contrast and blood flow noises presented in the images. 
In addition, the algorithm should be light-weight to be deployed for practical clinical applications. Therefore, it is desired to design a light-weight network with high performance for retinal layer segmentation. In this paper, we propose \mname~for retinal layer segmentation which can be applied to OCT images. Specifically, our approach follows an encoder-decoder structure, where the encoder part employs multi-scale feature extraction and a Transformer block for fully exploiting the semantic information of feature maps at all scales and making the features have better global reasoning capabilities, while the decoder part, we design a multi-scale asymmetric attention (MAA) module for preserving the semantic information at each encoder scale. The experiments show that our approach achieves a better segmentation performance compared to the current state-of-the-art method TransUnet with 105.7M parameters on both our collected dataset and two other public datasets, with only 3.3M parameters.
\end{abstract}

\begin{IEEEkeywords}
retinal layer segmentation, light-weight, multi-scale asymmetric attention, visible-light OCT.
\end{IEEEkeywords}

\section{INTRODUCTION}
\label{sec:INTRODUCTION}
With the accelerated pace of lifestyle and the popularity of electronic products, many people lack awareness of eye protection, leading to an increasing incidence of retinal diseases. Many retinal diseases are accompanied by changes in the thickness of the retinal layers. For instance, in patients with glaucoma, the retinal nerve fiber layer is thinner and the optic disc is atrophied compared to the healthy eyes \cite{horn2009correlation,pollet2014structure}. In diabetic retinopathy, the macular region of the retina thickens and becomes edematous~\cite{ajaz2021review}. 
Current research indicates that the changes in the retinal layer begin to occur in the early stages of the disease\cite{brandl2019retinal,van2019retinal}, therefore, automatic analysis of the retinal layers and monitoring their morphological changes can help to understand the disease progression and provide early treatment.

OCT can be used for non-invasive 3D structural imaging of the retina~\cite{1991Optical}, with which experienced ophthalmologists can manually inspect the retinal layers to identify the disease progression. However, manual inspection is inefficient and tedious. It is also challenging for various levels of ophthalmologists to ensure consistency and objectivity of the inspection. Although imaging techniques with a higher resolution are available now, such as Visible-light OCT images~\cite{shu2017visible,song2021wide,song2022linear}, this does not address completely the above status quo. Recent convolutional neural networks (CNNs), as typified by U-net~\cite{ronneberger2015u}, have achieved promising results regarding semantic segmentation in the field of medical image analysis. The seminar work ReLayNet~\cite{roy2017relaynet} has triggered widespread interest in applying CNNs to segment retinal layers. ReLayNet follows a multi-scale encoder-decoder structure, which has inspired many mainstream semantic segmentation frameworks to adopt, such as U-net~\cite{ronneberger2015u} and Attention\_Unet~\cite{oktay2018attention}. 
However, such structure can be limited in two main aspects. Firstly, most U-shape encoder-decoder structures perform the multi-scale feature extraction followed by feature fusion via only residual connections. Thus, it is limited to maintain and fully exploit the semantic information of feature maps at all scales, resulting in false positives in the background region when performing retinal layer segmentation. 
Secondly, many of these methods do not pay attention to the actual application needs of OCT devices, often pursuing accuracy unilaterally, for example, such multi-scale reasoning often comes at the cost of a large number of parameters, which can be computationally demanding for the deployment of real-time clinical applications. 

\begin{figure}[t]
\centering
\includegraphics[scale=0.55]{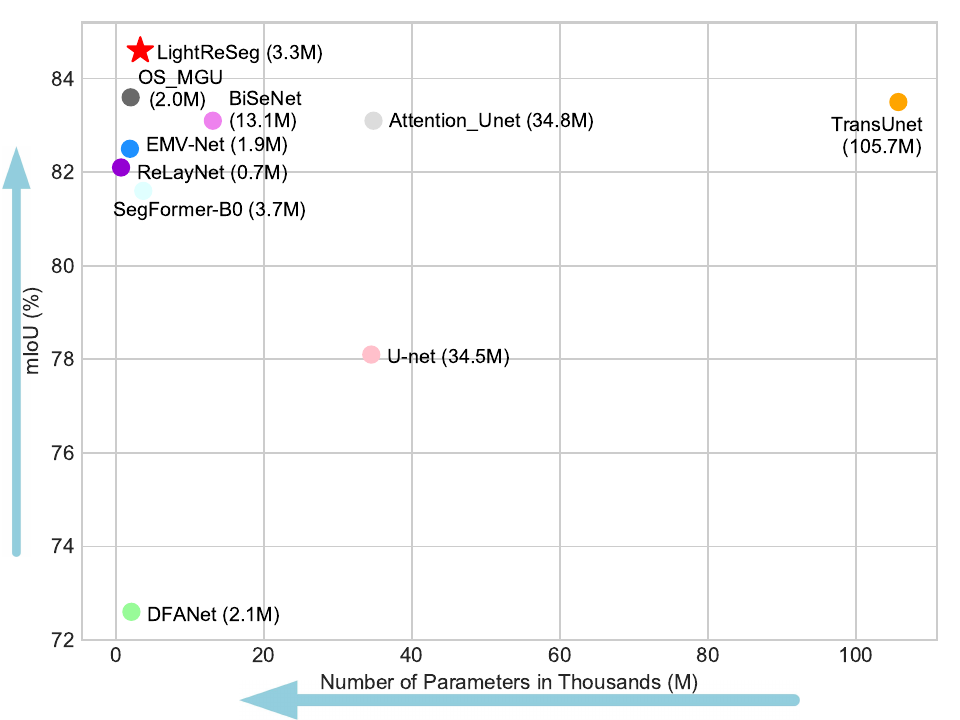}
\caption{The model size v.s. the segmentation accuracy in terms of mIoU of the state-of-the-art retinal layer segmentation methods. Our \mname~achieves the highest mIoU compared to SOTA methods while maintaining a smaller model size.}
\label{fig:mIoU_parameters}
\end{figure}

To overcome the above-mentioned limitations, we propose \mname, a novel multi-scale encoder-decoder network for end-to-end retinal layer segmentation, which is a light-weight segmentation approach tailored to practical OCT device requirements. 
In order to reduce segmentation errors in the background region, we propose to incorporate a Transformer block \cite{dosovitskiy2020image} to features with global reasoning and to introduce a multi-scale asymmetric attention (MAA) module to better preserve the semantic information at each encoder scale. 
To achieve computation efficiency, both the backbone and the MAA module adopt light-weight feature extractors, such as depthwise separable convolution and asymmetric convolution~\cite{chollet2017xception, ding2019acnet}. \mname~achieves the state-of-the-art segmentation accuracy with only 3.3M parameters, a significantly light model, as shown in Fig.~\ref{fig:mIoU_parameters}.

In summary, our contributions are: 
\begin{itemize}
    \item We propose \mname, a novel encoder-decoder structure for retinal layer segmentation, which exploits global information in a light-weight manner, to improve the segmentation accuracy with reduced computational complexity, it provides valuable experience for algorithms designed for practical applications in medical devices. 
    \item We propose a novel attention module, the MAA module, to jointly work with Transformer, in order to address erroneous segmentation in the background area by allowing the model to reason in a global manner.
    \item We perform our method on the visible-light OCT images for the first time and score the best segmentation performance, it provides experience for the algorithm to perform robustly on datasets from different domains.
\end{itemize}

This paper is organized as follows:
Sec.~\ref{sec:RELATED WORKS} describes the research progress in the field of biological tissue segmentation and retinal layer segmentation. 
Sec.~\ref{sec:METHODOLOGY} introduces the overall framework, the feature extractor, MAA module and the light-weight designs of the approach.
Sec.~\ref{sec:EXPERIMENT} describes the datasets, performance metrics, implementation details, analyzes quantitatively and qualitatively the experimental results, and performs ablation experiments.
Finally, the conclusions of the paper are given in Sec.~\ref{sec:CONCLUSION}.

\section{RELATED WORK}
\label{sec:RELATED WORKS}
In this section, we first provide an overview of the research progress of  semantic segmentation, especially light-weight semantic segmentation methods, followed by a more specific coverage of deep-learning-based retinal layer segmentation.

\subsection{Medical Image Segmentation}
\label{MIS}
Medical image segmentation has made rapid progress under the promotion of deep learning, especially in the fields of biological tissue recognition and lesion detection, which has sparked a large amount of research on medical image segmentation. 
The morphology of the optic disc and optic cup, as well as the cup-to-disc ratio, can be used to assess glaucoma, Wang et al. proposed an automatic segmentation approach based on CNNs to accurately segment the optic disc and optic cup from fundus images for glaucoma detection\cite{wang2019patch}. 
Precise segmentation of retinal vessels from fundus images is essential for intervention in numerous diseases and high-precision segmentation of retinal vessels still remains a challenging task, Li et al. proposed a dual-path progressive fusion network, named DPF-Net, which achieved better segmentation ability by effectively fusing global and local features\cite{li2023dpf}.
Abdominal multi-organ image segmentation plays a crucial role in the diagnosis and treatment of many diseases. traditional methods of manual depiction are inefficient and subjective, therefore, Chen et al. proposed TransUnet based on CNN and Transformer block for abdominal multi-organ segmentation\cite{chen2021transunet}. TransUnet was the first model to introduce the Transformer block into the U-shaped encoder-decoder structure and achieved very high accuracy in segmentation. 
There are also segmentation studies dedicated to lesion regions, such as Astaraki et al. who segmented different types of lung pathological regions to enable screening for lung diseases\cite{astaraki2022prior}. 
In addition, some lightweight segmentation methods for practical applications have also been proposed, such as SegFormer\cite{xie2021segformer}, A-net\cite{chen2023net}, EMV-Net\cite{he2023exploiting}.

\begin{figure*}[h]
\centering
\includegraphics[scale=0.8]{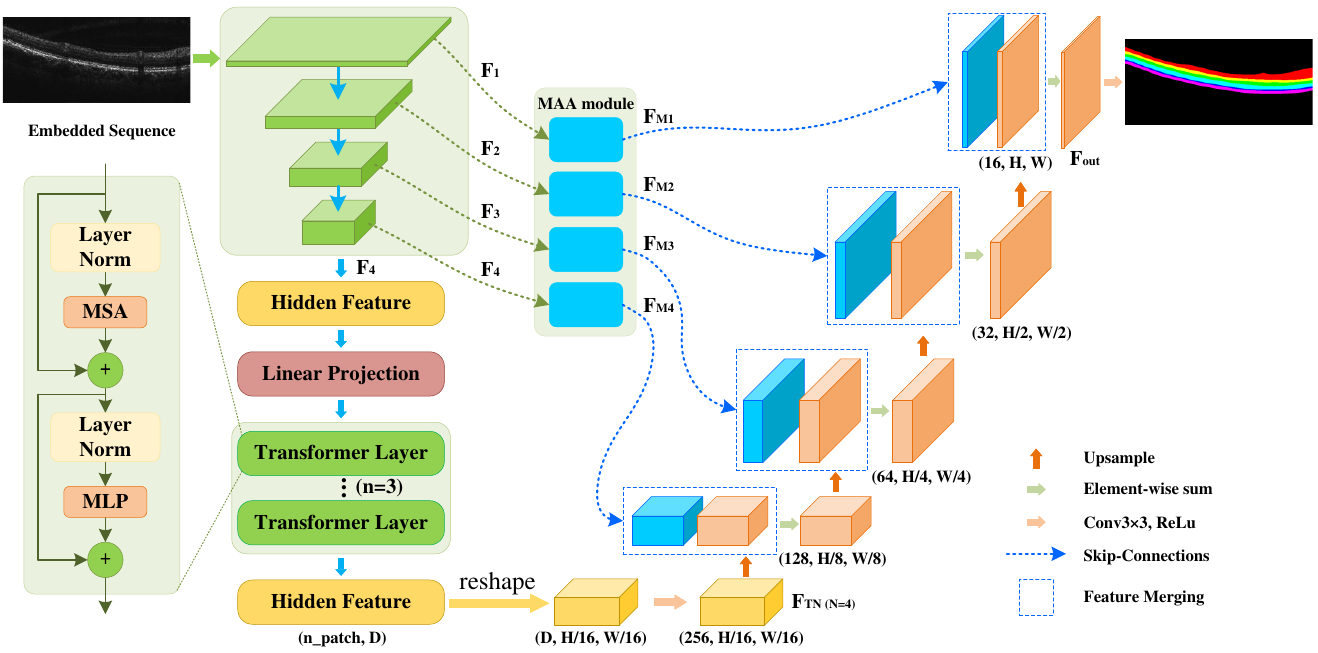}
\caption{The network of \mname~follows a U-shape encoder-decoder structure. The encoder takes as the input a retinal image of dimension $(3, H, W)$ and performs multi-scale feature extraction that outputs feature maps of $N$ scales, where $N$ is set to 4 in our design. The last feature map is further fed to the Transformer Layers through a linear transformation to extract features that reasons in long-range to help the reduction of the segmentation errors in the background region. 
The resulted features after the Transformer Layers are then fused via reshaping and up-sampling with the multi-scale encoder features that are optimized by the proposed MAA module. The final fused feature map $F_{out}$ is exploited for the retinal layer segmentation via convolutions.} 
\label{fig:whole approach}
\end{figure*}

\subsection{Retinal Layer Segmentation}
The morphology of the retinal layer is associated with the development of many ophthalmic diseases, therefore, there are many studies on retinal layer segmentation based on CNN and OCT images. The success of ReLayNet opened the door to the application of CNNs for retinal layer segmentation\cite{roy2017relaynet}, a typical U-shaped network based on an encoder-decoder structure, which achieved the best performance for retinal layer segmentation at that time. Since layer segmentation of OCT retinal images is prone to speckle noise, intensity inhomogeneity, Wang et al. proposed a boundary-aware U-Net for retinal layer segmentation by detecting accurate boundaries\cite{wang2021boundary}. 
To solve the topological errors caused by not considering the order of retinal layers, Liu et al proposed a novel deep learning-based framework that employed the distance maps of layer surfaces to convert the layer segmentation task into a multitasking problem for classification and regression\cite{liu2020confidence}.
To ensure the continuity of the retinal layer boundary and obtain more accurate segmentation results, Hu et al proposed a coarse-to-fine retinal layer boundary segmentation method based on the embedded residual recurrent network and the graph search, it has achieved good performance in both qualitative and quantitative indicators\cite{hu2021embedded}.
In addition, considering the impact of different neurological diseases on the retinal layer, Gende et al presented a fully automatic approach for the retinal layer segmentation in multiple neurodegenerative disorder scenarios\cite{gende2023automatic}, the results indicate that the model is more robust.
There are also studies aimed at practical application deployment, such as He et al. proposed a light-weight retinal layer segmentation approach that achieved good performance with a lower number of parameters\cite{he2023exploiting}.

\section{METHODOLOGY}
\label{sec:METHODOLOGY}

\subsection{Framework Overview}
Our proposed \mname~is a U-shape network based on an encoder-decoder structure as shown in Fig.~\ref{fig:whole approach}. 
The network takes a retinal layer OCT image as input. The first $N$ scale feature maps ($F_{1}$, ..., $F_{N-1}$, $F_{N}$) are extracted by the encoder which consists of a multi-scale feature extractor, respectively. In order to further deepen the global reasoning capability in the depth direction, the feature map $F_{N}$, which has to pass through a depth feature encoder: Transformer block\cite{dosovitskiy2020image}. The feature map $F_{TN}$ further encoded by the Transformer block has no change in shape and size, but it facilitates long-range global reasoning. To better preserve details at different scales, multi-scale features ($F_{1}$, ..., $F_{N-1}$, $F_{N}$) are then merged with its corresponding layers in the decoder part via our proposed MAA module, which outputs the feature maps ($F_{M1}$, ..., $F_{MN-1}$, $F_{MN}$) with the same shape size as its input. The $F_{TN}$ output from the Transformer block is merged with the $F_{MN}$ for feature fusion, and then after 2 fold size up-sampling, it continues to fuse with $F_{MN-1}$ for feature fusion and then follows a similar operation until the fusion of the up-sampled restored feature map with $F_{out}$ is completed, and after channel adjustment, the final retinal layer segmentation image is output. \mname~is designed for real-time clinical operations, where light-weight designs are employed where possible. 

We present in full the multi-scale encoder in Section~\ref{sec:method:encoder} followed by the proposed MAA module that serves for multi-scale feature fusion in Section~\ref{sec:method:MAA}. Finally, we describe the design details for computation efficiency in Section~\ref{sec:method:lightweight}.

\subsection{Multi-scale Encoder}
\label{sec:method:encoder}
The multi-scale encoder extracts $N$ scale features to fully exploit the semantic information of feature maps at all scales, further, in order to make the features have better global reasoning capabilities, we incorporate a Transformer block\cite{dosovitskiy2020image}, which is a module that further optimizes the $N$th scale feature map extracted from the deepest scale of the encoder. To ensure the dimensionality is consistent with the Transformer block port, we pre-process the feature map $F_N$ before being processed by the Transformer block.
The feature map $F_{N}$ is first converted into a $2D$ patch sequence as

\begin{equation}
\label{e1}
F_{N}\Rightarrow F_{N}^{'}=\left \{\right.x_{p}^{k}\in R^{P^{2}\times c}~|~k=1,...,Z\left. \right \}~,~Z=\frac{HW}{P^{2}},
\end{equation}
where $c$ is the number of channels in the $F_{N}$ feature map, $P$ is the size of the patch, and $H$ and $W$ are the height and width of the $F_{N}$ feature map. 

We use a trainable linear projection to map the vector patch $x_{p}$ to a latent D-dimensional embedding space. To encode the spatial information of the patches, we learn speciﬁc position embeddings that are added to preserve the positional information in the patch embedding as

\begin{equation}
\label{e2}
z_{0}=\left[x_{{0}} ; x_{p}^{1}E; x_{p}^{2}E; \cdots ; x_{p}^{N}E\right]+E_{pos},
\end{equation}
where $E \in \mathbb{R}^{\left(P^{2} \cdot C\right) \times D}$ represents the patch embedding projection, the $x_{0}$ is an additional learnable vector
concatenated with the remaining vectors to integrate the information of all remaining vectors, the $\left[\cdot;\cdots;\cdot\right]$ is the concatenation operator, and $E_{pos} \in \mathbb{R}^{(Z+1) \times D}$ represents the position embedding\cite{dosovitskiy2020image}.


\begin{figure*}[t]
\centering
\includegraphics[scale=0.85]{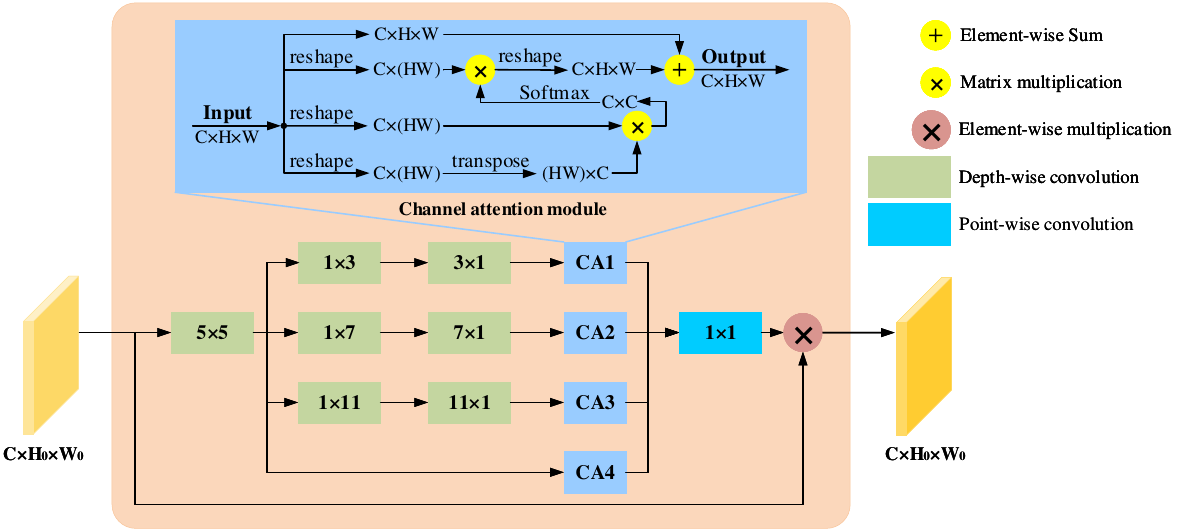}
\caption{The structure of Multi-scale Asymmetric Attention module.}
\label{MAA module}
\end{figure*}

Then, inputting $z_{0}$ into the Transformer Layer, as shown in Fig.~\ref{fig:whole approach}, first goes through the first Layer Norm (LN) layer and then enters the Multi-head Self-attention (MSA) layer, here $MSA\left(LN\left(z_{0}\right)\right)+z_{0}$ forms the residual structure and get the $z_{0}^{\prime}$. Then, after passing through the second LN layer, it enters the Multi-Layer Perceptron (MLP) layer, here $MSA\left(LN\left(z_{0}^{\prime}\right)\right)+z_{0}^{\prime}$ once again forms the residual structure and get the $z_{1}$.

At this point, the first Transformer layer is finished. We have set three Transformer layers in our approach, so we have to cycle three times.

\begin{equation}
\begin{split}
\label{e5}
z_{L}=\left[x_{{0}}^{'} ; x_{p}^{1'}; x_{p}^{2'}; \cdots ; x_{p}^{N'}\right]\Rightarrow \left[x_{p}^{1'}; x_{p}^{2'}; \cdots ; x_{p}^{N'}\right]
\end{split}
\end{equation}

When the $z_{L}$ outputs from the Transformer block, we have to remove the $x_{{0}}^{'}$ vector from the $z_{L}$, because only then we can reshape it to the same size of $F_{N}$.

\begin{figure*}[t]
\centering
\includegraphics[scale=0.95]{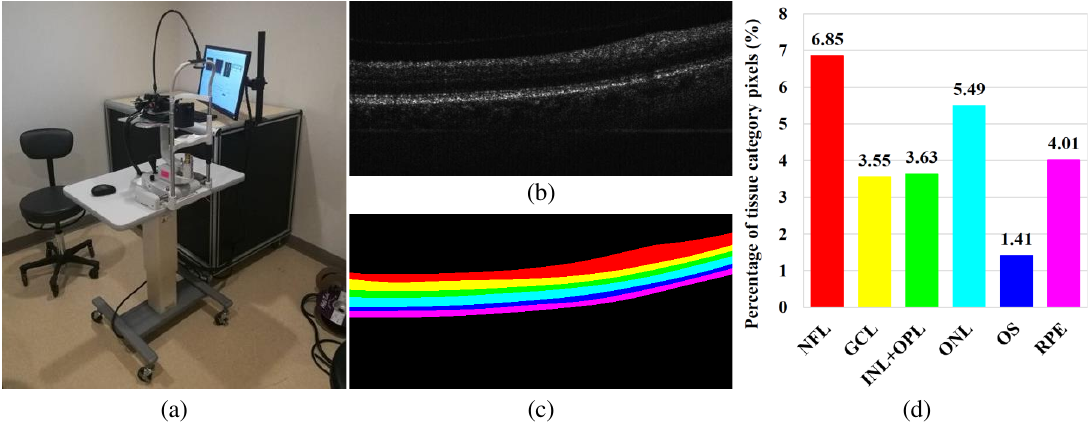}
\caption{(a) The visible light OCT device for capturing images. (b) Visible-light OCT B-scan image. (c) Annotation image (ground truth). (d) Average percentage of pixels on OCT images for each tissue layer besides the background(the percentage of background is 75.06\%).}
\label{fig:equipment}
\end{figure*}

\subsection{MAA Module}
\label{sec:method:MAA}
To better preserve the semantic information at each encoder scale, the MAA module is designed in the skip connection part of the encoder and decoder by a large convolution kernel and channel attention mechanism. It's also set with multi-scale feature fusion, and it's based on the ideas of asymmetric convolution\cite{ding2019acnet}, multi-scale feature extraction\cite{guo2022segnext,chen2017deeplab} and channel attention mechanism\cite{fu2019dual}. 
\begin{equation}
\label{e6}
F_{0}^{'}=Conv_{5\times 5}\left ( F_{0}\right )
\end{equation}
\begin{equation}
\label{e7}
F_{0i}^{'}=Conv_{ki\times 1}\left [Conv_{1\times ki}\left ( F_{0}^{'}\right )\right ], ~k=3, 7, 11
\end{equation}
\begin{equation}
\label{e8}
F_{0i}^{''}=CA\left ( F_{0i}^{'}\right ),
\end{equation}

The workflow of the MAA module is illustrated in Fig.~\ref{MAA module}, it begins with the input of a feature map $F_{0}$ extracted from the feature encoder, assuming a shape size of ($C$, $H_{0}$, $W_{0}$). $F_{0}$ enters MAA and goes through a $5\times 5$ convolution kernel to expand the perceptual field, and then has to enter a multi-scale convolution to extract information at different scales (Eq.~\ref{e6}, Eq.~\ref{e7}). Each scale goes through a channel attention (CA) module to increase the weight of the feature map on the channel dimension (Eq.~\ref{e8}). As shown in the channel attention module in Fig.~\ref{MAA module}, the feature $F_{0i}^{'}\in \mathbb{R}^{C\times H\times W}$ reshape to $F_{0i\_cn}^{'}\in\mathbb{R}^{C\times N}$, where $N = H\times W$, $\in$ represents the shape of the feature map. After that we perform a matrix multiplication between $F_{0i\_cn}^{'}$ and the transpose of it, get the $F_{0i\_cc}^{'}\in \mathbb{R}^{C\times C}$. Eventually, we apply the softmax layer to get the channel attention map $A\in\mathbb{R}^{C\times C}$ \begin{equation}
\label{e9}
F_{0i\_cc}^{'}=F_{0i\_cn}^{'} \times \left (F_{0i\_cn}^{'}\right )^{T}
\end{equation}
\begin{equation}
\label{e10}
A=Softmax\left (F_{0i\_cc}^{'}\right )
\end{equation}
(Eq.~\ref{e9}, Eq.~\ref{e10}). Here the matrix $A\in\mathbb{R}^{C\times C}$ contains information about the weights between channels, e.g. $\left (A\in\mathbb{R}^{C\times C}\right )_{ij}$ represents the impact of $i^{th}$ channel on $j^{th}$ channel. In addition, we multiply the matrix between the transpose of $A\in\mathbb{R}^{C\times C}$ and
\begin{equation}
\label{e11}
A_{1}\in\mathbb{R}^{C\times N}=A^{T}\times F_{0i\_cn}^{'}\\
\end{equation}
\begin{equation}
\label{e12}
F_{0i}^{''}=\alpha A_{1}\in\mathbb{R}^{C\times N\overset{reshape}{\rightarrow}C\times H\times W}+F_{0i}^{'}
\end{equation}
$F_{0i\_cn}^{'}$, and reshape the result to $\mathbb{R}^{C\times H\times W}$. Then we multiply the result by a scale parameter $\alpha $ and perform an element-wise sum operation with $F_{0i}^{'}$ to obtain the final output $F_{0i}^{''}$ (Eq.~\ref{e11} and Eq.~\ref{e12}). After the CA module, all feature maps are summed and convolved by $1\times 1$ kernel to generate a weighted feature map, and we use the weighted feature map to re-weight $F_{0}$ by element-wise multiplying ($\bigotimes$) it with the input $F_{0}$ (Eq.~\ref{e13} and Eq.~\ref{e14}).
\begin{equation}
\label{e13}
att=Conv_{1\times 1}\left [ \sum_{i=1}^{3}F_{0i}^{''}+CA\left ( F_{0}^{'}\right )\right ]
\end{equation}
\begin{equation}
\label{e14}
F_{M0}=att\bigotimes F_{0}
\end{equation}
At last, we obtain the $F_{M0}$. According to the above procedure, we can input the extracted four scales feature maps $F_{1}$, ..., $F_{N-1}$, $F_{N}$ into MAA module to get $F_{M1}$, ..., $F_{MN-1}$, $F_{MN}$, respectively, as shown in Fig.~\ref{fig:whole approach}. 

Instead, the ASPP employs a series of dilated convolution layers with different rates to obtain larger receptive fields, however, dilated convolutions may lose local detail information. In contrast, the MAA module utilizes asymmetric convolutions with large kernels, which allows capturing multi-scale features without sacrificing too much detail. Furthermore, before the 1x1 convolution compresses the channels in the MAA module, a channel attention mechanism is adopted, enhancing the module's focus on the channel dimension and better preserving important features.

\begin{table*}[b]
\centering
\caption{Multiple approaches to multiple metrics ($S_{mPA}$ (\%), $S_{mIoU}$ (\%), $S_{DSC}$ (\%), $S_{PA}$ (\%)) evaluation on the Vis-105H dataset, the No. 1 and No. 2 places are bolded in black and bolded in gray for each metric, respectively.}
\label{tab1}
\resizebox{0.85\textwidth}{!}{
\begin{tabular}{llcccccccc}
\hline
\hline
\multicolumn{1}{c}{\multirow{2}{*}{Method}} & \multirow{2}{*}{Indicators} & \multicolumn{6}{c}{Tissue Layers}          & \multirow{2}{*}{$S_{mPA}$} & \multirow{2}{*}{$S_{mIoU}$} \\ \cline{3-8}
\multicolumn{1}{c}{}                        &                             & NFL  & GCL  & INL/OPL & ONL  & OS   & RPE  &                        &                         \\ \hline
\multirow{2}{*}{ReLayNet\cite{roy2017relaynet}} & $S_{DSC}$               & 93.6 & 85.6 & 85.5    & 93.5 & 87.9 & 94.1 & \multirow{2}{*}{97.3}  & \multirow{2}{*}{82.1}   \\
                                            & $S_{PA}$                    & 93.3 & 82.1 & 86.0    & 97.2 & 87.5 & 91.1 &                        &                         \\ \cline{2-8}
\multirow{2}{*}{OS\_MGU\cite{Li:21}}        & $S_{DSC}$                   & \textcolor{gray}{\textbf{94.8}} & 86.8 & 86.7    & \textcolor{gray}{\textbf{94.1}} & \textcolor{gray}{\textbf{88.6}} & \textcolor{gray}{\textbf{94.5}} & \multirow{2}{*}{\textcolor{gray}{\textbf{97.7}}}  & \multirow{2}{*}{\textcolor{gray}{\textbf{83.6}}}   \\
                                            & $S_{PA}$                    & 93.5 & 82.8 & 87.4    & \textbf{97.8} & \textcolor{gray}{\textbf{88.1}} &91.2 &                        &                         \\ \cline{2-8}
\multirow{2}{*}{DFANet\cite{li2019dfanet}}  & $S_{DSC}$                   & 91.1 & 80.0 & 80.8    & 88.3 & 73.5 & 88.8 & \multirow{2}{*}{95.1}  & \multirow{2}{*}{72.6}   \\
                                            & $S_{PA}$                    & 91.3 & 77.0 & 82.9    & 93.3 & 74.4 & 86.3 &                        &                         \\ \cline{2-8}
\multirow{2}{*}{BiSeNet\cite{yu2018bisenet}}& $S_{DSC}$                   & \textbf{94.9} & 87.3 & \textcolor{gray}{\textbf{88.3}}    & 93.9 & 84.8 & \textcolor{gray}{\textbf{94.5}} & \multirow{2}{*}{\textbf{97.8}}  & \multirow{2}{*}{83.1}   \\
                                            & $S_{PA}$                    & \textbf{95.4} & \textcolor{gray}{\textbf{83.9}} & \textbf{89.7}    & 96.6 & 84.9 & 91.2 &                        &                         \\ \cline{2-8}
\multirow{2}{*}{Attention\_Unet\cite{oktay2018attention}}  & $S_{DSC}$    & 94.6 & \textcolor{gray}{\textbf{87.4}} & 87.6    & 93.6 & 87.1 & 93.7 & \multirow{2}{*}{97.6}  & \multirow{2}{*}{83.1}   \\
                                            & $S_{PA}$                    & \textcolor{gray}{\textbf{94.6}} & 83.4 & 87.6    & \textcolor{gray}{\textbf{97.6}} & 87.5 & 91.2 &                        &                         \\ \cline{2-8}
\multirow{2}{*}{EMV-Net\cite{he2023exploiting}}   & $S_{DSC}$           & 94.6 & 85.3 & 85.5    & 93.7 & 87.9 & 94.5 & \multirow{2}{*}{97.6}  & \multirow{2}{*}{82.5}   \\
                                            & $S_{PA}$                    & 93.5 & 81.9 & 86.2    & 97.1 &  \textbf{90.1} & \textbf{91.7} &                        &                         \\ \cline{2-8}
\multirow{2}{*}{U-net~\cite{ronneberger2015u}} & $S_{DSC}$             & 92.6 & 82.5 & 83.4    & 91.5 & 82.6 & 92.3 & \multirow{2}{*}{96.8}  & \multirow{2}{*}{78.1}   \\
                                            & $S_{PA}$                    & 90.3 & 80.8 & 84.8    & 96.6 & 80.8 & 90.0 &                        &                         \\ \cline{2-8}
\multirow{2}{*}{SegFormer-B0~\cite{xie2021segformer}} & $S_{DSC}$             & 94.5 & 83.4 & 84.1    & 93.4 & 88.2 & 94.2 & \multirow{2}{*}{97.4}  & \multirow{2}{*}{81.6}   \\
                                            & $S_{PA}$                    & 92.8 & 79.0 & 86.9    &  \textcolor{gray}{\textbf{97.6}} & 86.3 & 91.3 &                        &                         \\ \cline{2-8}
\multirow{2}{*}{TransUnet\cite{chen2021transunet}}  & $S_{DSC}$           & 94.4 & 86.9 & 87.5    & \textcolor{gray}{\textbf{94.1}} & 87.5 & \textbf{94.7} & \multirow{2}{*}{\textcolor{gray}{\textbf{97.7}}}  & \multirow{2}{*}{83.5}   \\
                                            & $S_{PA}$                    & 93.4 & 83.2 & 88.0    & 97.0 & 87.2 & \textcolor{gray}{\textbf{91.6}} &                        &                         \\ \cline{2-8}
\multirow{2}{*}{\mname~(Our)}               & $S_{DSC}$                   & 94.3 & \textbf{87.5} & \textbf{89.1}    & \textbf{94.7} & \textbf{89.1} & \textcolor{gray}{\textbf{94.5}} & \multirow{2}{*}{\textbf{97.8}}  & \multirow{2}{*}{\textbf{84.6}}   \\
                                            & $S_{PA}$                    & 94.3 & \textbf{84.5} & \textcolor{gray}{\textbf{89.4}}    & 96.9 & 87.8 & \textcolor{gray}{\textbf{91.6}} &                        &                         \\ \hline \hline
\end{tabular}
}
\end{table*}
\subsection{Light-weight Designs}
\label{sec:method:lightweight}

The basic principle of our design of \mname~is to improve the overall performance of the model with as few parameters as possible, so we can meet the computational efficiency of deployments for real-time clinical applications. Thus, we make some light-weight designs for part of the structure of the model. For encoder extraction of multi-scale feature maps, we use depthwise separable convolutions (DS-Conv) for our down-sampling\cite{chollet2017xception}. DS-Conv considers the channels and spatial regions of the feature map separately, takes different convolution kernels for different channels, and then uses point convolution to aggregate the channel information, achieving a richer feature representation with fewer parameters to learn. We find that using standard convolution for down-sampling is better than pooling, but convolution increases the number of parameters, so we use DS-Conv to ensure the segmentation performance while decreasing the burden of the number of parameters. In addition, as mentioned in Sec.~\ref{sec:method:MAA}, multi-scale asymmetric convolution is used in our MAA module. Initially, our idea is to perform multi-scale extraction of feature information like ASPP~\cite{chen2017deeplab}, but we find that using standard convolution would cause a greater burden on the number of parameters, since the MAA module is supposed to perform extraction of the $N$ scale feature maps, which would certainly further increase the number of parameters. In order to reduce the number of parameters in the MAA module, we use asymmetric convolution instead of standard convolution, shown in Fig.~\ref{MAA module}, and the overall performance of \mname~is nearly the same. In particular, the Transformer block has a significant proportion of the number of parameters in \mname. We have optimized the internal parameters of the Transformer block structure, such as the setting of heads=8 for multiple heads in MSA and dim\_head=64 for hidden linear layers in MSA\cite{dosovitskiy2020image}, etc. All these designs to minimize the number of parameters are made under the precondition of ensuring the performance of \mname.

\section{EXPERIMENT}
\label{sec:EXPERIMENT}
In this section, we compare our \mname~against the state-of-the-art methods for retinal layer segmentation on i) two publicly available OCT datasets with unhealthy eyes, where the Glaucoma dataset is for the Glaucoma disease and the DME dataset is for Diabetic Macular Edema (DME), and ii) a new dataset, named as Vis-105H, that we create on top of raw images collected from~\cite{song2022visible} which correspond to only healthy human eyes with visible light OCT. Moreover, we present a thorough ablation study with our Vis-105H dataset to validate the effectiveness of the proposed MAA and the introduction of the Transformer block. 

\subsection{Datasets}
\label{sec:datasets}

\paragraph{Vis-105H dataset}
We create the Vis-105H dataset for the method evaluation which contains images captured using a custom \textit{visible light} OCT device that takes optic disc cubes from each subject~\cite{song2021visible,yi2020systems,song2018fiber}, as shown in Fig.~\ref{fig:equipment}(a). The raw images are originally collected and presented in~\cite{song2022visible}, and they correspond to 21 eyes from 14 healthy subjects, all of whom with written informed consent prior to the participation. The inclusion criteria include i) age $40$ years, ii) best corrected visual acuity better than 20/40, iii) no previous intraocular surgery of any kind, and iv) no known retinal disease, resulting in 21 scanning cubes with 5376 B-scan images.
Two professional ophthalmologists further select 5 images with high imaging quality from every 256 B-scans, as shown in Fig.~\ref{fig:equipment}(b), with the selection criteria including i) low noise, ii) clear retinal layer boundaries, and iii) suitability for ophthalmologists to use for diagnosis. 
With the supervision of medical ophthalmology professionals, we annotate these 105 pre-processed retinal Visible-light OCT images of healthy human eyes for semantic segmentation, and complete the Vis-105H dataset for 7-class semantic segmentation including the background, as shown in Fig.~\ref{fig:equipment}(c). 
Specifically, we exploit the graphics software Inkscape to annotate the six layers, nerve fiber layer (NFL), ganglion cell layer (GCL) and inner plexiform layer (IPL), inner nuclear layer (INL) and out plexiform layer (OPL), outer nuclear layer (ONL), outer segment (OS), and retinal pigment epithelium (RPE), as red, yellow, green, light blue, dark blue, and pink, respectively, and to annotate the remaining regions as background black. Each annotated image is further examined by a professional ophthalmologist and exported in PNG format with a size of 660$\times$300 pixels. 
Fig.~\ref{fig:equipment}(d) shows the average pixel percentages of all categories on the OCT image except for the background, from which we can observe that the most dominant two classes are the NFL and ONL layers while the OS and GCL layers are the least dominant classes. The Vis-105H dataset is divided into a training set, a validation set, and a test set consisting of 75, 15, and 15 images, respectively. The data for the Vis-105 dataset is conducted at BMC, and both BU and BMC hold ownership rights to the data.

\paragraph{Glaucoma dataset} The dataset is collected from 61 different subjects~\cite{Li:21}, where 12 radial OCT B scans of each subject are collected using DRI OCT-1 Atlantis at the Ophthalmology Department of Shanghai General Hospital. All images are scanned at 20.48 mm × 7.94 mm field of view in the optic nerve head region. Under the supervision of a glaucoma specialist, two ophthalmologists manually annotate these images into the optic disc and nine retinal layers. The image size is 1024 × 992 and the dataset is divided into training, validation, and test sets by 148, 48, and 48 images, respectively. 

\paragraph{DME dataset} The dataset is collected by Chiu et al. using the Duke Enterprise Data Unified Content Explorer search engine to retrospectively identify DME subjects within the Duke Eye Center\cite{chiu2015kernel}. It consists of 110 OCT B-scans obtained from 10 patients with DME with a size of 496$\times$768 pixels, each B-scan is annotated with 9 layers. The dataset is divided by 88, 11, 11 into training, validation, and test set, respectively. All subjects are scanned with the macula at the center.

\begin{table*}[t]
\centering
\caption{Multiple approaches to multiple metrics ($S_{mPA}$ (\%), $S_{mIoU}$ (\%), $S_{DSC}$ (\%), $S_{PA}$ (\%)) evaluation on the DME dataset, the No. 1 and No. 2 places are bolded in black and bolded in gray for each metric, respectively.}
\label{tab2}
\resizebox{0.9\textwidth}{!}{
\begin{tabular}{llcccccccccc}
\hline
\hline
\multicolumn{1}{c}{\multirow{2}{*}{Method}} & \multirow{2}{*}{Indicators} & \multicolumn{8}{c}{Tissue Layers}                                                                                          & \multirow{2}{*}{$S_{mPA}$} & \multirow{2}{*}{$S_{mIoU}$} \\ \cline{3-10}
\multicolumn{1}{c}{}                        &                             & \multicolumn{1}{c}{NFL}  & GCL/IPL & \multicolumn{1}{c}{INL}  & \multicolumn{1}{c}{OPL}  & ONL/ISM & ISE  & OS/RPE & Fluid &                        &                         \\ \hline
\multirow{2}{*}{ReLayNet\cite{roy2017relaynet}}                   & $S_{DSC}$                       & 82.0                     & 93.4    & 79.4                     & 77.2                     & 86.8    & 86.6 & \textcolor{gray}{\textbf{86.5}}   & 53.8  & \multirow{2}{*}{95.5}  & \multirow{2}{*}{69.6}   \\
                                            & $S_{PA}$                       & 79.5                     & 93.3    & 76.2                     & 76.7                     & 87.7    &89.7 & 87.2   & 52.0  &                        &                         \\ \cline{2-10}
\multirow{2}{*}{OS\_MGU\cite{Li:21}}                     & $S_{DSC}$                       & \textbf{82.6}                     & \textcolor{gray}{\textbf{93.6}}    & \textcolor{gray}{\textbf{79.5}}                     & \textcolor{gray}{\textbf{78.4}}                     & 87.2    & 87.1 & 85.6   & 57.7  & \multirow{2}{*}{\textcolor{gray}{\textbf{95.6}}}  & \multirow{2}{*}{69.8}   \\
                                            & $S_{PA}$                       & \textcolor{gray}{\textbf{80.7}}                     &  \textcolor{gray}{\textbf{93.7}}    & 77.0                     & 77.2                     &\textcolor{gray}{\textbf{88.3}}    & 89.5 & 84.6   & 52.1  &                        &                         \\ \cline{2-10}
\multirow{2}{*}{DFANet\cite{li2019dfanet}}                     & $S_{DSC}$                       & 77.4                     & 90.3    & 73.6                     & 72.6                     & 85.8    & 85.7 & 85.5   & 49.8  & \multirow{2}{*}{94.7}  & \multirow{2}{*}{64.8}   \\
                                            & $S_{PA}$                       & 73.7                     & 91.4    & 70.4                     & 70.8                     & 84.1    & 89.4 & 85.9   & 47.3  &                        &                         \\ \cline{2-10}
\multirow{2}{*}{BiSeNet\cite{yu2018bisenet}}                    & $S_{DSC}$                       & 80.9                     & 92.7    & 77.9                     & 75.3                     & 86.3    & 86.1 & \textbf{86.6}   & 56.7  & \multirow{2}{*}{95.4}  & \multirow{2}{*}{68.3}   \\
                                            & $S_{PA}$                       & 76.7                     & 93.0    & 76.8                     & 72.2                     & 86.9    & 87.5 &  \textbf{89.8} & 51.0  &                        &                         \\ \cline{2-10}
\multirow{2}{*}{Attention\_Unet\cite{oktay2018attention}}            & $S_{DSC}$                       & 80.5                     & 91.5    & 77.6                     & 75.9                     & 86.8    & 86.6 & 86.4   & 55.6  & \multirow{2}{*}{95.3}  & \multirow{2}{*}{68.0}   \\
                                            & $S_{PA}$                       & 79.4                     & 92.4    & 74.4                     & 74.0                     & 86.8    & 88.9 & 86.6   & 49.9  &                        &                         \\ \cline{2-10}
\multirow{2}{*}{EMV-Net\cite{he2023exploiting}}                 & $S_{DSC}$                       & 80.4                     & 92.5    & 78.5                     & 74.8                    & 85.6   & 86.3 & 86.3   & \textcolor{gray}{\textbf{59.9}}  & \multirow{2}{*}{95.3}  & \multirow{2}{*}{68.4}   \\
                                            & $S_{PA}$                       & 74.9 & 91.6    &  \textbf{81.8} & 71.3 & 81.2    & 89.4 &87.6   & \textbf{60.2}  &                        &                         \\ \cline{2-10}
\multirow{2}{*}{U-net~\cite{ronneberger2015u}}                      & $S_{DSC}$                       & 82.2                     & 92.8    & 79.1                     & 78.2                     & 86.7    & \textbf{87.4} & \textbf{86.6}   & 56.2  & \multirow{2}{*}{95.5}  & \multirow{2}{*}{69.4}   \\
                                            & $S_{PA}$                       &\textbf{81.4} & 92.9    &76.2 &76.5 &87.6    & \textbf{90.7} &86.7   &52.2  &                        &                         \\ \cline{2-10}
\multirow{2}{*}{SegFormer-B0~\cite{xie2021segformer}}                      & $S_{DSC}$                       & 81.7                     & \textbf{93.7}    & 79.1             & 77.0              & 86.6    & 86.6 & 85.7   & 57.6  & \multirow{2}{*}{95.6}  & \multirow{2}{*}{69.2}   \\
                                            & $S_{PA}$                       &77.5 &  \textbf{95.0}    &74.9 &77.0 & \textbf{89.8}    & \textcolor{gray}{\textbf{89.9}} & \textcolor{gray}{\textbf{87.8}}   &49.1  &                        &                         \\ \cline{2-10}
\multirow{2}{*}{TransUnet\cite{chen2021transunet}}                  & $S_{DSC}$                       & 81.4                     & 93.1    & \textbf{79.8}                     & 77.4                     & \textcolor{gray}{\textbf{87.4}}    &87.2 & \textbf{86.6}   & \textbf{60.1}  & \multirow{2}{*}{\textbf{95.7}}  & \multirow{2}{*}{\textcolor{gray}{\textbf{70.0}}}   \\
                                            & $S_{PA}$                       &80.0 & 93.0    & \textcolor{gray}{\textbf{77.7}} & \textcolor{gray}{\textbf{77.6}} & 88.2    & 89.5 & 86.7   &53.9  &                        &                         \\ \cline{2-10}
\multirow{2}{*}{\mname~(Our)}                & $S_{DSC}$                       & \textcolor{gray}{\textbf{82.4}}                     & \textbf{93.7}    & \textbf{79.8}                     & \textbf{78.8}                     & \textbf{87.9}    &  \textcolor{gray}{\textbf{87.3}} & 86.0   & 59.7  & \multirow{2}{*}{\textbf{95.7}}  & \multirow{2}{*}{\textbf{70.5}}   \\
                                            & $S_{PA}$                       & \multicolumn{1}{c}{79.7} & 93.5    & \multicolumn{1}{c}{77.1} & \multicolumn{1}{l}{\textbf{78.1}} & 87.8    & 89.6 & 85.6   &  \textcolor{gray}{\textbf{57.8}}  &                        &                         \\ \hline \hline
\end{tabular}
}
\end{table*}
\begin{table*}[b]
\centering
\caption{Multiple approaches to multiple metrics ($S_{mPA}$ (\%), $S_{mIoU}$ (\%), $S_{DSC}$ (\%), $S_{PA}$ (\%)) evaluation on the Glaucoma dataset, the No. 1 and No. 2 places are bolded in black and bolded in gray for each metric, respectively.}
\label{tab3}
\resizebox{0.9\textwidth}{!}{
\begin{tabular}{llcccccccccccc}
\hline
\hline
\multicolumn{1}{c}{\multirow{2}{*}{Method}} & \multirow{2}{*}{Indicators} & \multicolumn{10}{c}{Tissue Layers}                                                                                                                                                              & \multirow{2}{*}{$S_{mPA}$} & \multirow{2}{*}{$S_{mIoU}$} \\ \cline{3-12}
\multicolumn{1}{c}{}                        &                             & \multicolumn{1}{c}{NFL}  & \multicolumn{1}{c}{GCL}  & \multicolumn{1}{c}{IPL}  & \multicolumn{1}{l}{INL}  & \multicolumn{1}{l}{OPL}  & \multicolumn{1}{c}{ONL}  & IS/OS & RPE  & Choroid & OD   &                        &                         \\ \hline
\multirow{2}{*}{ReLayNet\cite{roy2017relaynet}}                   & $S_{DSC}$    & 79.4   & \textcolor{gray}{\textbf{64.9}}         & \textbf{70.8}       & \textcolor{gray}{\textbf{76.4}}                     & \textbf{81.1}                     & 90.5                     & \textcolor{gray}{\textbf{86.0}}    & \textbf{93.1} & 87.9    & 77.8 & \multirow{2}{*}{94.2}  & \multirow{2}{*}{\textcolor{gray}{\textbf{67.0}}}   \\
                                            & $S_{PA}$                       & 77.2                     & \textcolor{gray}{\textbf{60.7}}    & \textcolor{gray}{\textbf{70.6}}     & 77.0    & \textbf{80.5}                     & 90.0                     & 84.9  & 85.2 & 88.6    & 77.4 &                        &                         \\ \cline{2-12}
\multirow{2}{*}{OS-MGU\cite{Li:21}}     & $S_{DSC}$      &80.7    & 62.0   & 69.6      & 74.2    & 78.4    & 89.6   & 84.7  & \textcolor{gray}{\textbf{83.0}} &88.5    & 83.7 & \multirow{2}{*}{\textcolor{gray}{\textbf{94.8}}}  & \multirow{2}{*}{66.7}   \\
                                            & $S_{PA}$                       & 82.8                     & 56.1                     & 68.7                     & 74.4                     & 75.7                     & 88.3                     & 85.0  & \textbf{85.7} & 88.7    & 86.8 &                        &                         \\ \cline{2-12}
\multirow{2}{*}{DFANet\cite{li2019dfanet}}                     & $S_{DSC}$   & 79.8     & 60.5                     & 68.2                     & 74.3                     & 77.4                     & 89.4                     & 85.0  & 81.3 & 88.2    &83.9 & \multirow{2}{*}{94.7}  & \multirow{2}{*}{65.8}   \\
                                            & $S_{PA}$                       & 80.0                     & 55.9                     & 66.1                     & 71.9                     & 73.1                     & 91.7                     & \textbf{85.7}  & 80.8 & 89.6    & 88.6 &                        &                         \\ \cline{2-12}
\multirow{2}{*}{BiSeNet\cite{yu2018bisenet}}                    & $S_{DSC}$                       & 78.0                     & 61.5                     & 67.4                     & 70.1                     & 77.1                     & 88.5                     & 82.4  & 78.3 & 86.9    & \textbf{84.3} & \multirow{2}{*}{94.5}  & \multirow{2}{*}{64.0}   \\
                                            & $S_{PA}$                       & 76.4                     & 57.0                     & 65.6                     & 66.2                     & 77.4                     & 89.5                     & 82.3  & 81.6 &90.1    & \textcolor{gray}{\textbf{89.3}} &                        &                         \\ \cline{2-12}
\multirow{2}{*}{Attention\_Unet\cite{oktay2018attention}}            & $S_{DSC}$       & 79.5  & 60.1    & 64.7   & 72.8   & 78.8                     & \textcolor{gray}{\textbf{90.6}}                     & \textbf{86.3}  & 82.8 & \textbf{88.8}    & 83.2 & \multirow{2}{*}{94.7}  & \multirow{2}{*}{66.0}   \\
                                            & $S_{PA}$                       & 84.2     & 57.9     & 59.4    & 71.0      & 75.0                     & 91.9                    & \textcolor{gray}{\textbf{85.6}}  & 81.8 & 89.1    & 84.8 &                        &                         \\ \cline{2-12}
\multirow{2}{*}{EMV-Net\cite{he2023exploiting}}                 & $S_{DSC}$                       & 81.2                     & \textbf{65.3}                    & 69.5                     & 72.1                     & 79.1                     & 90.0                     &83.9  & 81.5 & 88.5    & 83.4 & \multirow{2}{*}{\textcolor{gray}{\textbf{94.8}}}  & \multirow{2}{*}{66.6}   \\
                                            & $S_{PA}$                       &83.7 & 59.3 & 67.8 & 65.8 & 74.8 & 88.6 & 81.5  & 83.4 & 89.9    & 86.8 &                        &                         \\ \cline{2-12}
\multirow{2}{*}{U-net~\cite{ronneberger2015u}}      & $S_{DSC}$         &  \textcolor{gray}{\textbf{81.9}}             & 63.7     & 69.4         & 73.5           & 77.5           & 89.7          & 85.4  & 82.3 & 87.8    & 83.3 & \multirow{2}{*}{\textcolor{gray}{\textbf{94.8}}}  & \multirow{2}{*}{66.6}   \\
                                            & $S_{PA}$                       & 83.5 & 58.1 & 68.3 & 73.7 & 74.6& 91.7 & 84.2  & 83.4 &\textbf{91.0}    & 86.5 &                        &                         \\ \cline{2-12}
\multirow{2}{*}{SegFormer-B0~\cite{xie2021segformer}}    & $S_{DSC}$     & \textbf{82.1}    & 64.3    & 69.4 & \textcolor{gray}{\textbf{76.4}}       & 76.6   & 89.9    & 84.4  & 81.2 &  \textcolor{gray}{\textbf{88.6}}    &  \textcolor{gray}{\textbf{84.2}} & \multirow{2}{*}{\textcolor{gray}{\textbf{94.8}}}  & \multirow{2}{*}{66.9}   \\
                                            & $S_{PA}$                       & \textbf{86.7} & 58.5 & 63.9 & \textbf{81.7} & 69.7& \textcolor{gray}{\textbf{92.6}} & 82.8  & 79.8&87.7    &  \textbf{90.7} &                        &                         \\ \cline{2-12}
\multirow{2}{*}{TransUnet\cite{chen2021transunet}}                  & $S_{DSC}$      & 80.6       & 60.0       & 69.4         & 76.1      & 80.2        & \textbf{90.8}        & 85.8  & 82.6 &88.5    & 82.4 & \multirow{2}{*}{94.7}  & \multirow{2}{*}{\textcolor{gray}{\textbf{67.0}}}   \\
                                            & $S_{PA}$                       &\textcolor{gray}{\textbf{84.7}} & 53.3 & 69.0 & 77.6 & 78.5 &\textbf{93.6} & 83.0  & 80.3 & 86.9    & 84.3 &                        &                         \\ \cline{2-12}
\multirow{2}{*}{\mname~(Our)}     & $S_{DSC}$    & 80.9   &64.3   & \textcolor{gray}{\textbf{70.5}}   & \textbf{77.2}   & \textcolor{gray}{\textbf{80.4}}  & 90.1    & 85.4  & 82.7 & 88.4    & 83.3 & \multirow{2}{*}{\textbf{94.9}}  & \multirow{2}{*}{\textbf{67.8}}   \\
                                            & $S_{PA}$                       & \multicolumn{1}{c}{83.5} & \multicolumn{1}{c}{\textbf{61.1}} & \multicolumn{1}{c}{\textbf{70.8}} & \multicolumn{1}{l}{\textcolor{gray}{\textbf{78.8}}} & \multicolumn{1}{l}{\textcolor{gray}{\textbf{78.7}}} & \multicolumn{1}{c}{90.2} & 85.0  & \textcolor{gray}{\textbf{85.3}} & \textcolor{gray}{\textbf{90.5}}    & 83.5 &                        &                         \\ \hline \hline
\end{tabular}
}
\end{table*}
\subsection{Performance Metrics}
We evaluate the segmentation performance quantitatively using the Dice similarity coefficient (DSC), mean Intersection over Union (mIoU), pixel accuracy (PA), and mean pixel accuracy (mPA). They can be computed as:
\begin{equation}
\label{dsc}
S_{DSC} = \frac{2\times TP}{2\times TP+FP+FN},
\end{equation}
\begin{equation}
\label{mIoU}
S_{mIoU} = \frac{1}{k+1} \sum_{i=0}^{k} \frac{TP}{FN+FP+TP},
\end{equation}
\begin{equation}
\label{PA}
S_{PA} = \frac{TP+TN}{TP+FP+TN+FN} 
\end{equation}
and
\begin{equation}
\label{mPA}
S_{mPA} = \frac{1}{k+1} \sum_{i=0}^{k} \frac{TP+TN}{TP+FP+TN+FN},
\end{equation}
where $TP$, $FP$, $TN$, and $FN$ represent True Positives, False Positives, True Negatives, and False Negatives respectively. $k$ represents the number of categories. In addition, we count the number of parameters for compared approaches to represent their computation complexity.

\subsection{Implementation Details}
\mname~is implemented based on the PyTorch and trained with the Adam optimizer with the cross-entropy loss. The initial learning rate is set to 0.001 which is then gradually halved every 40 epochs. Data augmentation is applied on all three datasets, including horizontal flipping with probability P=0.5, random center rotation within plus or minus 20 degrees with probability P=0.5, median and motion blur processing with P=0.5 probability, random Gaussian noise addition, as well as random brightness and contrast with P=0.5 probability. All experiments are conducted on an NVIDIA GeForce RTX 3090 Graphics card. The code of the proposed \mname~could be found at: https://github.com/Medical-Image-Analysis/LightReSeg.

\subsection{Comparison}
\label{sec:COMPARISON AND DISCUSSION}

We compare \mname~with the state-of-the-art approaches including ReLaynet\cite{roy2017relaynet}, EMV-Net\cite{he2023exploiting}, Attention\_Unet\cite{oktay2018attention}, BiSeNet\cite{yu2018bisenet}, DFANet\cite{li2019dfanet}, OS\_MGU\cite{Li:21}, U-net~\cite{ronneberger2015u} and TransUnet\cite{chen2021transunet}. We report the results on the above-mentioned three datasets.

\subsubsection{Quantitative Analysis}
\label{Quantitative Analysis}
Tab.~\ref{tab1} shows the comparison of our approach with state-of-the-art methods on the Vis-105H dataset. 
We can see that our approach \mname~scores the best performance in both $S_{mPA}$ and $S_{mIoU}$ metrics, with $+1\%$ improvement in terms of the $S_{mIoU}$ compared to the second-best performing method OS\_MGU~\cite{Li:21}. 
In terms of the $S_{DSC}$ metric, our approach achieves the highest segmentation performance in the GCL layer, INL/OPL layer, ONL layer, and OS layer. Regarding the $S_{PA}$ metric, our approach achieves the first place in the GCL layer, and the second place in the INL/OPL layer and RPE layer, which also shows the outstanding performance of our proposed approach. 
The segmentation performance of ReLayNet, the lightest algorithm in the field of retinal layer segmentation, is $-2.5\%$ lower than our proposed \mname~in terms of the $S_{mIoU}$ metric. 
We further perform a statistical significance test, using the Wilcoxon rank sum test, to compare the Dice Score performance of different methods on each layer. When comparing our approach with ReLayNet, we observe a $P_{value}$ of 0.031250 (p$<$0.05), indicating a statistically significant difference.
The poor performance of ReLayNet might be due to its rather simple feature fusion strategy for combining features from encoder and decoder, i.e. skip connections, leading to limited semantic information at the fusion. Instead, \mname~further introduces the MAA module for the feature fusion at multiple scales. Moreover, we add Transformer layers at the bottom of the encoder for efficient global reasoning which further improves the segmentation accuracy. 
SegFormer is a new light-weight model in the field of semantic segmentation and has a good performance in many tasks~\cite{xie2021segformer}, but as can be seen from Tab.~\ref{tab1}, SegFormer-B0 only ranks 7th in terms of the $S_{mIoU}$ and $S_{mPA}$ metrics, while in terms of the $S_{DSC}$ metric, its segmentation performance of all layers is much lower than ours. In terms of the $S_{mPA}$ metric, only the ONL layer is higher than our by 0.7 percentage points, and the segmentation performance in the rest of the layers is much lower than our approach.
We analyze that the poor segmentation performance of SegFormer-B0 on this task may be mainly due to two factors: first, its excessive pursuit of lightweight results in limited ability to extract detailed features; second, the up-sampling part of SegFormer-B0's decoder is too rough, resulting in the lack of fine detail recovery.

\begin{figure}[h]
\centering
\includegraphics[scale=0.55]{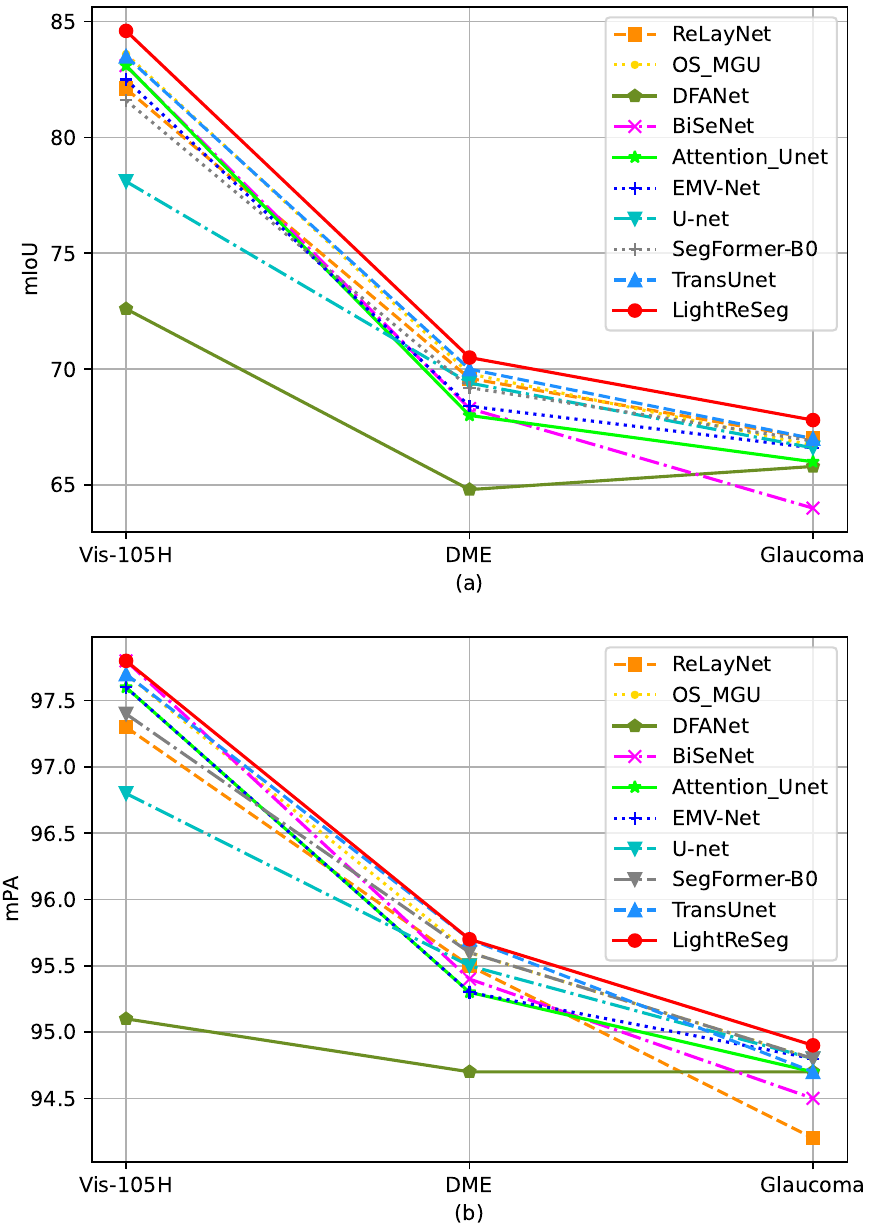}
\caption{(a) and (b) are the performance of the nine mainstream approaches on the three retinal layer segmentation datasets measured by the mIoU and mPA metrics, respectively.}
\label{fig:mPA_mIoU}
\end{figure}

We evaluate our approach on the DME dataset, as shown in Tab.~\ref{tab2}. We achieve the best performance in terms of both $S_{mPA}$ and $S_{mIoU}$. 
In terms of $S_{DSC}$ we achieve first place in 4 layers (i.e. GCL/IPL, INL, OPL, and ONL/ISM) and second place in another 2 layers (i.e. NFL and ISE), while in terms of $S_{mPA}$, we are on par with the state-of-the-art performance. 
Among other approaches, such as TransUnet, OS\_MGU, and ReLayNet, although they are more advanced in the $S_{mPA}$ metric, they are 0.5, 0.7, and 0.9 percentage points lower than the first place in the $S_{mIoU}$ metric, respectively, which also indicates that our model also has a strong comprehensive performance on the DME dataset. 
We further perform the statistical significance test by using the Wilcoxon rank sum test. For example, when comparing our method with TransUnet on the same domain, we observe a $P_{value}$ of 0.017629 (p$<$0.05), indicating a statistically significant difference. Similar statistically significant differences are observed, with $P_{values}$ of 0.023437 (p$<$0.05), 0.039062 (p$<$0.05), and 0.017755 (p$<$0.05) respectively when comparing our method with OS\_MGU, EMV-Net, and SegFormer-B0.
We further evaluate our approach on the Glaucoma dataset, as shown in Tab.~\ref{tab3}. Our \mname~achieves the best results on both the overall metrics $S_{mPA}$ and $S_{mIoU}$. Additionally, our approach also scores better on most of the layers in terms of both $S_{DSC}$ and $S_{PA}$. 

Fig.~\ref{fig:mPA_mIoU} (a) and (b) show the performance of the ten mainstream approaches on the three datasets measured by the mIoU and mPA metrics, respectively.
Our approach \mname~achieve the best performance on all three datasets in terms of $S_{mIoU}$ and $S_{mPA}$. TransUnet performs is the second best-performing method with a small margin lower than ours on all three datasets, however, its number of parameters is 32 times of the one of our approach. The lightest model ReLayNet is inferior to our method in terms of performance by $-2.5\%$, $-0.9\%$, and $-0.8\%$ on the Vis-105H, DME, and Glaucoma dataset, respectively. Therefore, our approach achieves state-of-the-art performance with a relatively light-weight model. 

In addition, we find that regardless of the method, certain layers (NFL, ONL, RPE) have significantly higher accuracy compared to other layers, as shown in Tab.~\ref{tab1}. We believe that class imbalance is the most likely cause, as the number of samples in a certain class is much larger than in other classes, resulting in the model learning better for that class during training and exhibiting higher accuracy in evaluation. From Fig.~\ref{fig:equipment}(d), we can see that the average pixel proportion of NFL, ONL, and RPE layers is relatively high, and Tab.~\ref{tab1} also shows that these three layers have higher accuracy compared to other layers. Also, there are differences in shape, color, texture, and other aspects among different layers, which may make certain categories of targets easier to segment while others are more difficult. In summary, we believe that the best way to solve this problem is to first address class imbalance.

\begin{figure}[t]
\centering
\includegraphics[scale=0.75]{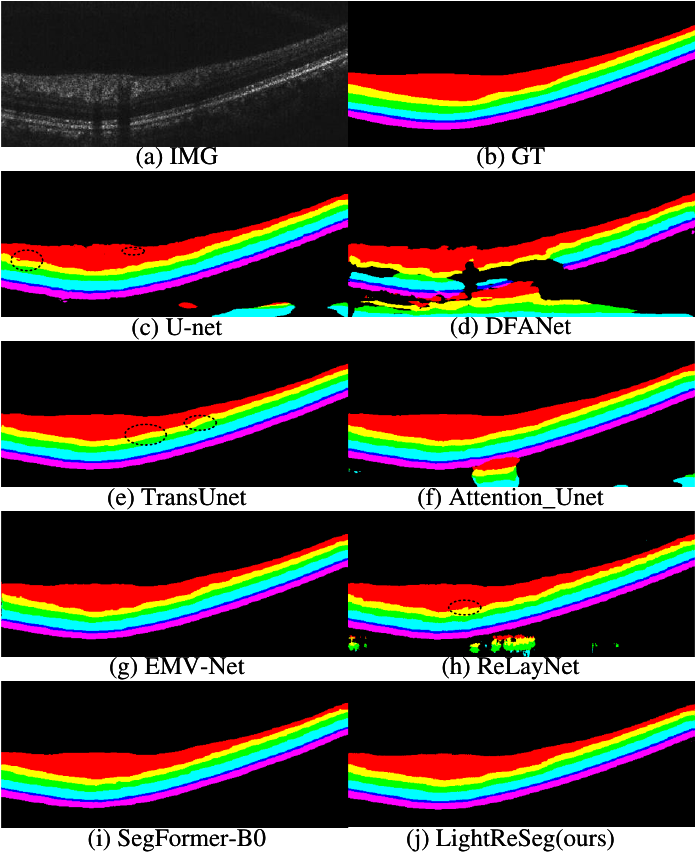}
\caption{Comparison of segmentation prediction maps of mainstream approaches on the Vis-105H dataset. (a) Original image. (b) Ground truth (c) Prediction map of U-net. (d) Prediction map of DFANet. (e) Prediction map of TransUnet. (f) Prediction map of Attention\_Unet. (g) Prediction map of EMV-Net. (h) Prediction map of ReLayNet. (i) Prediction map of SegFormer-B0. (j) Prediction map of LightReSeg. The black dashed line framed area in the image is the region where the boundary of the prediction layer is not smooth or inaccurate. }
\label{fig:Vis-105H}
\end{figure}

\begin{figure}[]
\centering
\includegraphics[scale=0.7]{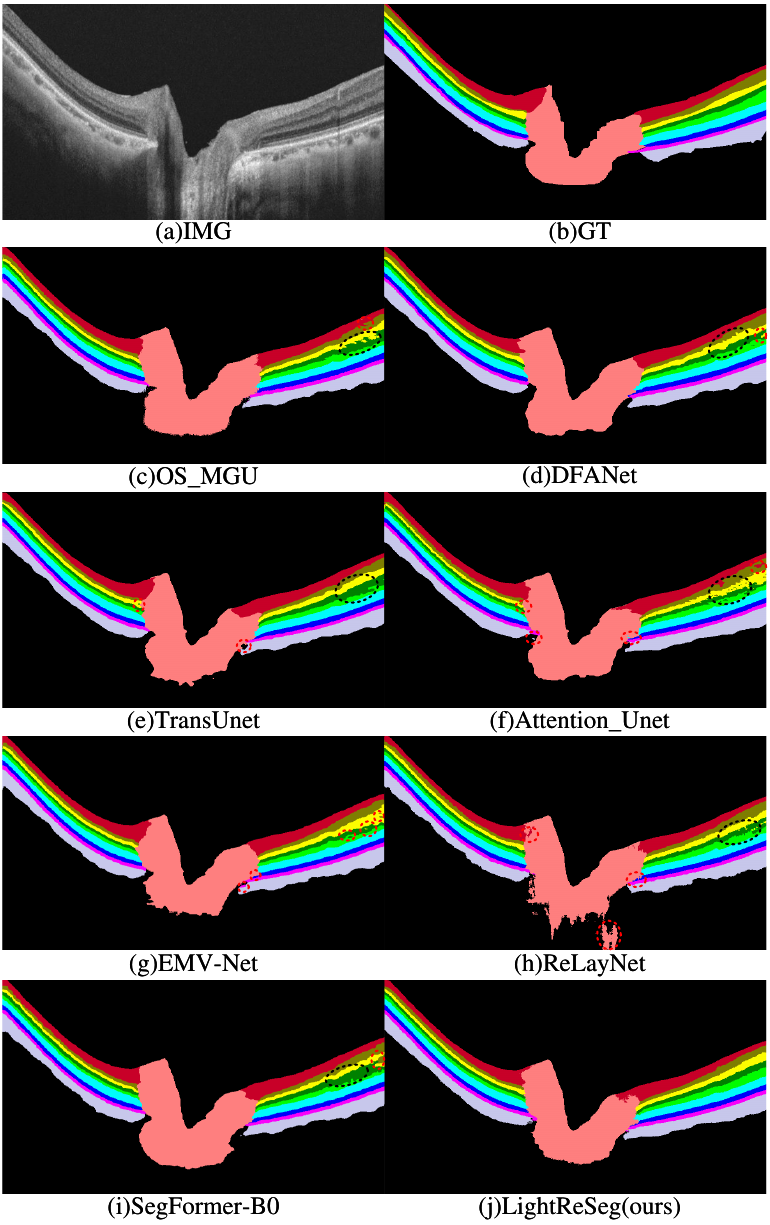}
\caption{Comparison of segmentation prediction maps of mainstream approaches on the Glaucoma dataset. (a) Original image. (b) Ground truth (c) Prediction map of OS\_MGU. (d) Prediction map of DFANet. (e) Prediction map of TransUnet. (f) Prediction map of Attention\_Unet. (g) Prediction map of EMV-Net. (h) Prediction map of ReLayNet. (i) Prediction map of SegFormer-B0. (j) Prediction map of LightReSeg. The black dashed line framed area in the image is the region where the boundary of the prediction layer is not smooth or inaccurate, the red dashed area is where the prediction map has a segmentation category error.}
\label{fig:Glaucoma}
\end{figure}

\subsubsection{Qualitative Analysis}
\label{sec:Qualitative Analysis}
Fig.~\ref{fig:Vis-105H} shows the segmentation prediction maps of several mainstream segmentation methods in a retinal B-scan image that contains blood flow information disturbance. As shown in Fig.~\ref{fig:Vis-105H}(j), our approach achieves much better segmentation performance. The prediction graphs of all the approaches in Fig.~\ref{fig:Vis-105H}(c), (d), (f), (h) all showed false positive errors, specifically, the region above the NFL layer or below the RPE layer of the retina is incorrectly identified as the retinal layer and segmented, resulting in the wrong segmentation target. The TransUnet in Fig.~\ref{fig:Vis-105H}(e) performs well, but two locations on the boundary between the NFL and GCL layers are inaccurately segmented, as marked by the two locations in Fig.~\ref{fig:Vis-105H}(e). The same occurs in EMV-Net in one place, as shown in Fig.~\ref{fig:Vis-105H}(g). In comparison, our approach is not only without false positive errors, but also segmentation boundaries are more accurate.

Fig.~\ref{fig:Glaucoma} shows the comparative prediction plots of several mainstream segmentation approaches in a single B-scan image over the retinal optic nerve head region. As shown in Fig.~\ref{fig:Glaucoma}(j), our approach shows a better segmentation performance. The black dashed region in Fig.~\ref{fig:Glaucoma}(d), where the NFL and GCL layer boundaries appear distinctly jagged, and the black dashed region in Fig.~\ref{fig:Glaucoma}(f), where a distinct honeycomb shape appears, are signs of unsmooth and inaccurate layer boundary prediction, and the same occurs in Fig.~\ref{fig:Glaucoma}(c), (e), and (h). In addition, the problem of segmentation category error occurs, such as the red dashed area in Fig.~\ref{fig:Glaucoma}(d), where the segmentation results in incorrectly segmenting part of the OPL layer into ONL layers, and this segmentation category error occurs at least twice in Fig.~\ref{fig:Glaucoma}, (e), (f), and (g). In contrast, our approach not only has no problem with segmentation category error but also the segmented boundary is smoother than other approaches.

\subsection{Ablation study}
\label{sec:exp:ablation}

\subsubsection{Performance of Different Modules}
\label{Different Modules}

\begin{figure}[h]
\centering
\includegraphics[scale=0.95]{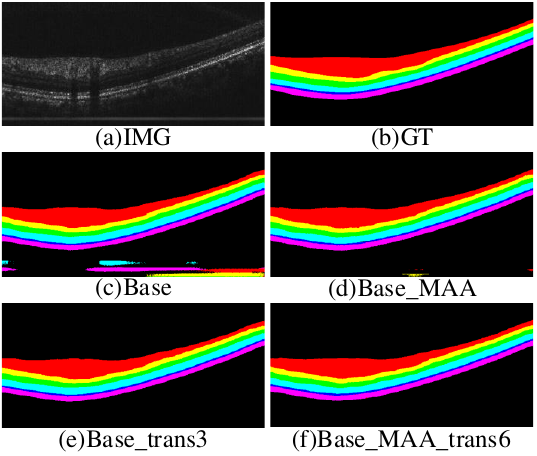}
\caption{Prediction images of our method under different ablation settings. (a) Original image. (b) Ground truth. (c) Base framework (d) Base framework plus MAA module (e) Base framework plus three Transformer blocks (d) Base framework plus MAA module and six Transformer blocks.}
\label{fig:Ablation}
\end{figure}

We conduct ablation experiments on the Vis-105H dataset with the main purpose of verifying the effect of the MAA module and the contribution of the Transformer block on the overall method segmentation performance. The ablated methods are: Base is the U-shaped backbone; Base\_MAA and Base\_trans3 are MAA and 3 transformer layers added to Base, respectively; Base\_MAA\_trans3 and Base\_MAA\_trans6 are based on Base\_MAA with 3 and 6 transformer layers respectively. Note that Base\_MAA\_trans3 represents the final version of our proposed \mname.
As shown in Tab.~\ref{ablation_vis}, the Base\_MAA approach improves over the Base approach, the $S_{mPA}$ and $S_{mIoU}$ by $+0.4\%$ and $+1.9\%$, respectively. All six layers except for the OS layer are also significantly improved in the $S_{DSC}$. We also observe that Base\_MAA\_trans3 improves over Base\_trans3 in terms of both $S_{mPA}$ and $S_{mIoU}$. Fig.~\ref{fig:Ablation}(c) and (d) show that the segmentation performance of the model is greatly improved by adding the MAA module to the model, and the results of mis-segmentation are significantly reduced in the background region below the RPE layer. 

Regarding the Transformer block, as shown in Tab.~\ref{ablation_vis}, Base\_trans3 improves $+0.5\%$ and $+2.4\%$ over Base in terms of $S_{mPA}$ and $S_{mIoU}$, respectively, and the segmentation ability of each layer is improved in terms of $S_{DSC}$. In addition, we also find that increasing the number of Transformer layers can further improve the segmentation performance within a certain range. For example, Base\_MAA\_trans6 adds 3 more Transformer layers than Base\_MAA\_trans3, which improves $S_{mPA}$ and $S_{mIoU}$ by $+0.1\%$ and $0.6\%$, respectively, this comes at a cost of $+40\%$ more of parameters. Fig.~\ref{fig:Ablation}(c) and (e) show qualitatively that adding three sequential Transformer blocks to the model contributes to a big improvement in the segmentation performance of the model, and there is no segmentation error in the background region below the RPE layer. 

\begin{table*}[t]
\centering
\caption{The proposed approach performs multiple metrics evaluation of ablation experiments on the Vis-105H dataset.}
\label{ablation_vis}
\resizebox{0.9\textwidth}{!}{
\begin{tabular}{lccccccccc}
\hline
\hline
\multicolumn{1}{c}{\multirow{2}{*}{Method}}                       & \multicolumn{6}{c}{$S_{DSC}$ (\%) Results of Tissue Layers}          &\multicolumn{1}{c}{\multirow{2}{*}{\begin{tabular}[c]{@{}c@{}}$S_{mPA}$\\ (\%)\end{tabular}}}&\multicolumn{1}{c}{\multirow{2}{*}{\begin{tabular}[c]{@{}c@{}}$S_{mIoU}$\\ (\%)\end{tabular}}} & \multirow{2}{*}{\begin{tabular}[c]{@{}c@{}}Params\\ (M)\end{tabular}} \\ \cline{2-7}
\multicolumn{1}{c}{}                                              & NFL  & GCL  & INL/OPL & ONL  & OS   & RPE  &                       &                        &                                                                       \\ \hline
Base                                                              & 94.0 & 84.6 & 84.8    & 93.2 & 87.9 & 92.6 & 97.2                  & 81.3                   & 1.61                                                                  \\ \cline{2-10} 
Base\_MAA                                                         & 94.4 & 86.9 & 87.7    & 94.0 & 87.4 & 94.0 & 97.6                  & 83.2                   & 1.72                                                                  \\ \cline{2-10} 
Base\_trans3                                                      & 94.7 & 86.9 & 87.3    & 94.1 & 89.0 & 94.0 & 97.7                  & 83.7                   & 3.20                                                                  \\ \cline{2-10} 
Base\_MAA\_trans3(Our)                      & 94.3 & 87.5 & 89.1    & 94.7 & 89.1 & 94.5 & 97.8                  & 84.6                   & 3.30                                                                  \\ \cline{2-10} 
Base\_MAA\_trans6                           & 95.0 & 87.6 & 89.0    & 94.8 & 89.7 & 95.1 & 97.9                  & 85.2                   & 4.68                                                                  \\ \hline \hline
\end{tabular}
}
\end{table*}

\begin{figure*}[t]
\centering
\includegraphics[scale=0.7]{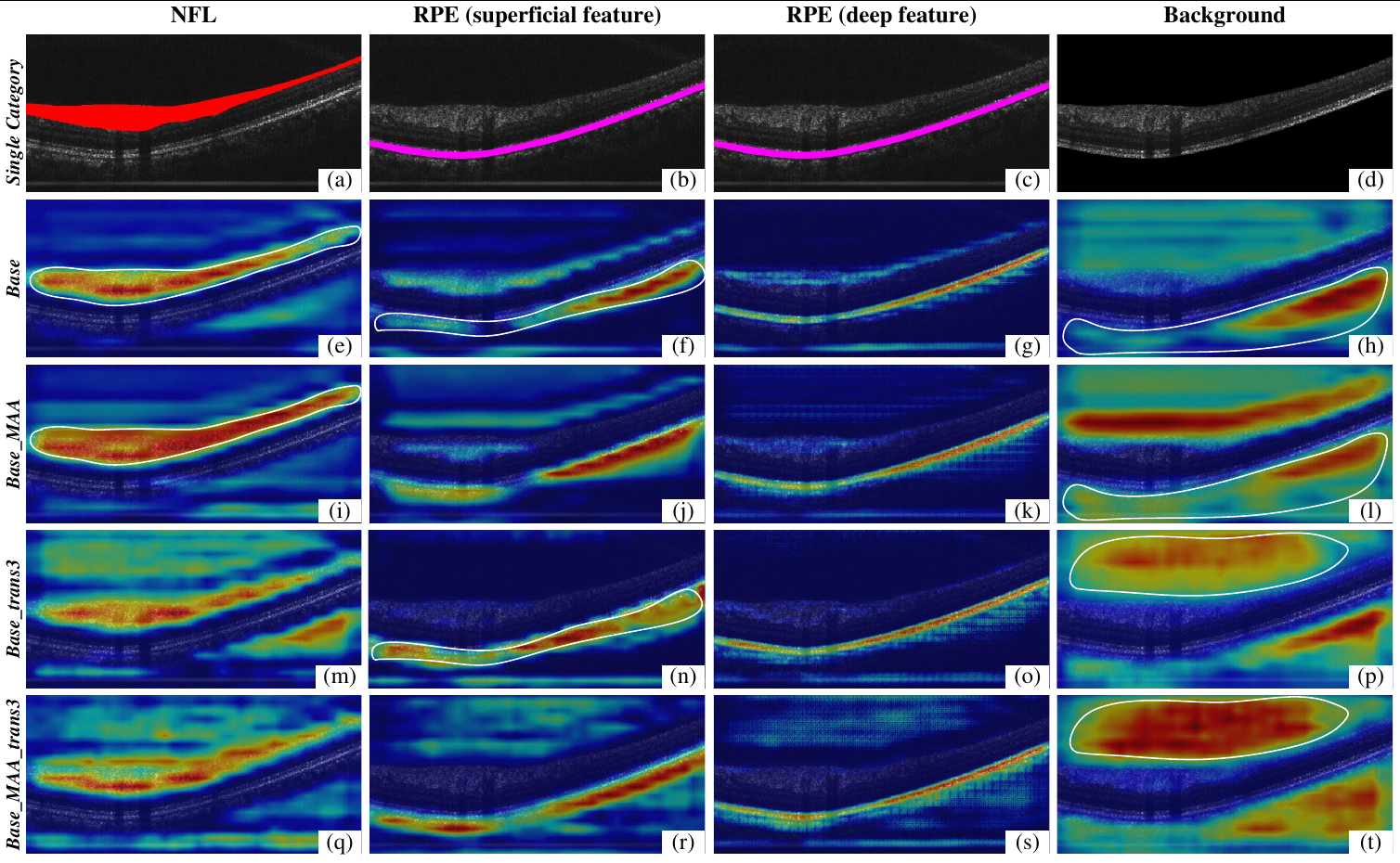}
\caption{Heatmap for different retinal layers at different model structures and at different positions of the model, (a–d) for the single category prediction, (e–h) for the Base structure, (i–l) for the Base\_MAA structure, (m–p) for the Base\_trans3 structure, and (q–t) for the Base\_MAA\_trans3 (Our) structure. The white line highlights the contrast area in the graph.}
\label{fig:Heatmap}
\end{figure*}

To further interpret the role of different modules, we use a modified version of Grad-CAM\cite{vinogradova2020towards} to visualize the feature activations in different layers of the model (see Fig.~\ref{fig:Heatmap}). Specifically, in accordance with the model settings in Tab.~\ref{ablation_vis}, we extract the feature activation heat maps for different retinal layers at different model structures and at different positions of the model, respectively. 
By comparing Fig.~\ref{fig:Heatmap}(e) and (i), with the addition of the MAA module, we can see that the heat map of the corresponding NFL layer in Fig.~\ref{fig:Heatmap}(i) is further enhanced compared to Fig.~\ref{fig:Heatmap}(e). The same trend is also observed in Fig.~\ref{fig:Heatmap}(h) and (l). 
It is further found that with the addition of the MAA module to the Base\_trans3 model structure, the feature region corresponding to the segmentation category is also further enhanced, as shown in Fig.~\ref{fig:Heatmap}(p) and (t). 
Adding the Transformer blocks to the model also causes changes in the heat map, as in Fig.~\ref{fig:Heatmap}(f) and (n). The heat map of the corresponding RPE layer region is activated more comprehensively with the addition of the Transformer blocks, and the OS and background layers of the adjacent regions also show different degrees of activation. 

When analyzing the activation maps, we observe that not only the regions corresponding to the predicted layers prominently highlighted, but also the feature weights of the adjacent layers are also significantly amplified. To interpret this behavior, we conduct further analysis and find that the boundaries of the retinal layers exhibit a high degree of correlation with the adjacent layers. As in Fig.~\ref{fig:Heatmap}(e), the focus is mainly on the NFL layer, but adjacent layers (e.g., background and GCL) also show various degrees of activation. Similarly, in Fig.~\ref{fig:Heatmap}(r), the RPE layer of the retinal layer is the main focus, while information from the OS and background layers at the upper and lower boundaries of the retinal layer is also utilized. This phenomenon suggests that the model uses information from adjacent layers to enhance its predictions. In addition, we also export the feature activation maps for different deep network layers in the prediction of the RPE layer, and we find that the heat map of the deep network focuses on more detailed and accurate regions compared to the heat map of the superficial network. As in Fig.~\ref{fig:Heatmap}(o), the regions where the heat map is focused on activation are highly coincident with the RPE layer, while the regions activated by the heat map in Fig.~\ref{fig:Heatmap}(n) are much coarser, which is similarly shown in Fig.~\ref{fig:Heatmap}(r) and (s). This observation suggests that the deeper extract image features of the model network contain more accurate and insightful semantic information. Overall, the visualization of deep feature activation provides valuable insights into the model's decision process, and the model can utilize multiple layers of information to obtain more robust segmentation results.

\subsubsection{Light-weight Setup}
\label{Light-weight Setup}

To evaluate the potential impact of DS-Conv on feature extraction capabilities, we replace all DS-Conv in the encoder section with standard convolutions and conduct comparative analyses\cite{howard2017mobilenets}, as shown in Tab.~\ref{DS-Conv}. The results indicate that using standard convolutions instead of DS-Conv does not enhance feature extraction in our optimized model and leads to an 11\% increase in the parameter count. This demonstrates that a simple increase in the model's parameter count does not always result in performance.
\begin{table}[]
\centering
\caption{Impact of DS-Conv on Model Performance.}
\label{DS-Conv}
\resizebox{0.45\textwidth}{!}{
\begin{tabular}{llll}
\hline
\hline
Method Settings              & $S_{mPA}$(\%) &$S_{mIoU}$(\%) & Params(M) \\ \hline
Encoder with DS-Conv & 97.8    & 84.6     & 3.30      \\ 
Encoder with Conv    & 97.8    & 84.3     & 3.69      \\ \hline
\hline
\end{tabular}
}
\end{table}

For assessing the impact of model light-weight on performance, we modify the number of channels in the encoder's multi-scale features, both halving and doubling them, the specific results are shown in Tab.~\ref{2N}. We find that, based on our model, when the channel of the multi-scale features extracted by the encoder is reduced to 0.5 and 0.25 times, the model's parameter count decreased, but its segmentation performance significantly deteriorated. When we increase the channel count to 2 and 4 times, there is no significant change in the model's accuracy, yet the parameter counts are 3 and 11 times the original size, respectively. From this, we can also conclude that our lightweight design maintains accuracy while using the fewest possible parameters.
\begin{table}[]
\centering
\caption{Impact of Model Light-weight on Performance.}
\label{2N}
\resizebox{0.45\textwidth}{!}{
\begin{tabular}{llll}
\hline
\hline
Method Settings & mPA(\%) & mIoU(\%) & Params(M) \\ \hline
0.25x           & 96.9    & 79.2     & 0.48      \\ 
0.5x             & 96.9    & 79.1     & 1.18      \\ 
1x               & 97.8    & 84.6     & 3.30      \\ 
2x               & 97.8    & 84.3     & 9.87      \\ 
4x               & 97.7    & 84.1     & 33.48     \\ \hline
\hline
\end{tabular}
}
\end{table}

\subsection{Limitation}
\label{limitation}
Although \mname~boasts only 3.3M trainable parameters and delivers the optimal retinal layer segmentation outcomes, in real-world applications, model parameter size is not the sole determinant of algorithm efficacy. Inference speed emerges as a crucial metric for assessing algorithmic effectiveness. As evidenced by Tab.~\ref{tab_limitation}, our proposed method exhibits inference speeds of 0.11$s$, 0.27$s$, and 0.07$s$ on the three datasets, respectively. While these figures already signify remarkable efficiency, there remains scope for further enhancement. For example, although the segmentation accuracy of ReLayNet is far inferior to our approach, it must be admitted that its inference efficiency is slightly higher than ours.
\begin{table}[h]
\centering
\caption{Statistical inference speed of different methods on three datasets, and the best performance on each dataset is bolded.}
\label{tab_limitation}
\resizebox{0.5\textwidth}{!}{
\begin{tabular}{lcccc}
\hline
\hline
Method  & \multicolumn{1}{c}{\begin{tabular}[c]{@{}c@{}}Params\end{tabular}} & \begin{tabular}[c]{@{}c@{}}Vis-105H\end{tabular} & \begin{tabular}[c]{@{}c@{}} DME\end{tabular} & \begin{tabular}[c]{@{}c@{}} Glaucoma\end{tabular} \\ \hline
{ReLayNet\cite{roy2017relaynet}}           & 0.7$M$    & 0.08$s$                    & 0.20$s$               & 0.06$s$                    \\
{EMV-Net\cite{he2023exploiting}}         & 1.9$M$   & 0.15$s$                    & 0.31$s$               & 0.13$s$                    \\
{OS\_MGU\cite{Li:21}}                     & 2.0$M$    & 0.12$s$                    & 0.30$s$               & 0.07$s$                    \\
{DFANet\cite{li2019dfanet}}                & 2.1$M$    & 0.12$s$                    & 0.31$s$               & 0.08$s$                    \\
{SegFormer-B0~\cite{xie2021segformer}}         &3.7$M$   & 0.12$s$                    & 0.28$s$               & 0.07$s$                    \\
{BiSeNet\cite{yu2018bisenet}}              & 13.1$M$   & 0.08$s$      & 0.20$s$               & 0.05$s$                    \\
{U-net~\cite{ronneberger2015u}}         & 34.5$M$   & 0.10$s$                    & 0.20$s$               & 0.08$s$                    \\
{Attention\_Unet\cite{oktay2018attention}} & 34.8$M$   & 0.10$s$                    & 0.22$s$               & 0.09$s$                    \\
{TransUnet\cite{chen2021transunet}}        & 105.7$M$  & 0.14$s$                    & 0.32$s$             & 0.11$s$                    \\
\mname~(Our)                            & 3.3$M$    & 0.11$s$       & 0.27$s$        & 0.07$s$                    \\ \hline \hline
\end{tabular}
}
\end{table}
Moreover, in the context of practical clinical applications, constraints related to dataset limitations preclude the inclusion of all device data types. Imagery obtained through disparate devices may exhibit domain discrepancies, inevitably resulting in false positive errors in the segmentation output for unfamiliar new devices. Consequently, apart from enlarging the dataset, enhancing the model's domain generalization competency is a subject worthy of further investigation. The supplementary segmentation result evaluation system will also augment ophthalmologists' confidence in the segmentation outcomes.

\section{CONCLUSION}
\label{sec:CONCLUSION}
In this paper, we propose a novel light-weight method \mname~for retinal layer segmentation. Our method introduces a Transformer-based block in the encoder part to enable global reasoning for reducing errors in the background region and an attention mechanism, named MAA module, to best exploit rich semantic information for the multi-scale feature fusion to improve the segmentation accuracy. The extensive evaluation shows that our approach achieves the best segmentation performance on both our collected Vis-105H dataset and two other public ones, i.e. DME and Glaucoma, indicating that our model has good reliability in the face of noise and uncertainty in the data. 
To improve the efficiency, our method also incorporates light-weight designs. The ablation experiments detailed in Section~\ref{Light-weight Setup} demonstrate that our lightweight design is precisely adequate, maintaining the highest level of accuracy while significantly reducing the number of parameters compared to other high-performing methods. This further highlights the practicality of our proposed approach, particularly evident in the real-time preview functionality of small-field-of-view OCTA imaging.
In the future, we will collect more clinical data to further improve our approach and continuously contribute to the efficient performance of OCT devices.

\bibliographystyle{IEEEtran} 
\bibliography{reference.bib} 

\end{document}